%% file: ms.tex
\documentclass[twocolumn,tradiabstract]{aa} 

\usepackage{natbib}
\usepackage{graphicx}

\usepackage{xcolor}

\begin{document}

\title{Numerical simulations of quiet Sun magnetic fields seeded by Biermann battery}
\titlerunning{Quiet Sun magnetic fields seeded by Biermann battery}

\author{E. Khomenko\inst{1,2}, N. Vitas\inst{1,2}, M. Collados\inst{1,2}, A. de Vicente\inst{1,2}}
\authorrunning{E. Khomenko et al.}

\institute{Instituto de Astrof\'{\i}sica de Canarias, 38205 La Laguna, Tenerife, Spain
\and Departamento de Astrof\'{\i}sica, Universidad de La Laguna, 38205, La Laguna, Tenerife, Spain}

\date{Received; Accepted }

\abstract {The magnetic fields of the quiet Sun cover at any time more than 90\% of its surface and their magnetic energy budget is crucial to explain the thermal structure of the solar atmosphere. One of the possible origins of these fields is due to the action of local dynamo in the upper convection zone of the Sun. Existing simulations of the local solar dynamo require an initial seed field, and sufficiently high spatial resolution, in order to achieve the amplification of the seed field to the observed values in the quiet Sun. Here we report an alternative model of seeding based on the action of the Bierman battery effect.  This effect generates a magnetic field due to the local imbalances in electron pressure in the partially ionized solar plasma. We show that the battery effect self-consistently creates from zero an initial seed field of a strength of the order of micro G, and  together with dynamo amplification, allows the generation of quiet Sun magnetic fields of a similar strength to those from solar observations.}


\maketitle

\section{Introduction} %

The solar magnetic field is known to be comprised of a large scale organized component manifest in the form of sunspots and active regions, and by another weaker component that exists in the quiet internetwork regions.
The quiet Sun component has a monotonic distribution of magnetic field strength with tails extending to kG values, and a complex distribution of inclinations  \citep{SanchezAlmeida+MartinezGonzales2011, deWijn2009, MartinezPillet2013, Danilovic2016}. Since polarimetric signals measured in the internetwork are at the limit of sensitivity of state-of-the-art instrumentation, a vivid discussion exists in the literature about the structure and mean magnetic field strength derived from observations in those regions. Observations with different resolutions and polarimetric sensitivities, and using spectral lines sensitive to Zeeman and Hanle effects, do not give the same results. While the analysis based on the Zeeman effect has the obvious drawback of the blindness to fields of mixed polarities on small spatial scales, the analysis based on the Hanle effect is dependent on models and diagnostic means, is limited by saturation effects, and requires assumptions on the turbulent nature of the field
\citep[see, however,][who tried to reduce this problem using differential Hanle measurements]{MansoSainz2004, Kleint2011}. Earlier Zeeman-based studies revealed values of the mean unsigned field strength at the base of the photosphere of the order of 20 G or less \citep{Khomenko+etal2005, BelloGonzalez2009}, later corrected to values of about 130--170 G \citep{Danilovic2010b, Danilovic2016}.
Similarly, first Hanle-based measurement provided rather low values \citep{Faurobert1993, Faurobert2001}, but the current understanding is that the mean value of the magnetic field strength in the quiet Sun is about 130 G \citep[this value refers to heights of a few hundred km above the photospheric base, see][]{TrujilloBueno2004, Shchukina+TrujilloBueno2011}. Lower mean values of  a few tens of G were also reported from observations in molecular lines \citep{Berdyugina+Fluri2004, Shapiro2007}, probably corresponding to granular regions. Even more controversy exists about the spatial structuring of this weak internetwork component \citep[e.g.,][]{Centeno2007, Lites2008, MartinezGonzalez2012, AsensioRamos+MartinezGonzalez2014, Lagg+etal2016}.

Given the uncertainty in the observational properties, the origin of the weak internetwork component has not been ultimately clarified neither \citep{Solanki2009}. There are hints that the properties of the quiet Sun fields are  independent of the solar cycle \citep{TrujilloBueno2004,  SanchezAlmeida2004, Buehler2013, Lites2014}, while \citet{Kleint2010} claimed to observe a very small variation of the turbulent field strength between the solar maximum and minimum. It has been proposed that the quiet Sun magnetic fields are produced by the local dynamo, acting in solar subsurface layers on granular scales \citep{Petrovay+Szakaly1993, Cattaneo1999}. This theory has been checked with the help of realistic numerical simulations of solar convection with state-of-the-art treatment of the relevant physics and improved treatment of numerical diffusivities and boundary conditions \citep{Vogler+Schussler2007, PietarilaGraham2010, Rempel2014, Kitiashvili2015} showing the possibility of local dynamo action amplifying an initial arbitrarily introduced seed of 10$^{-6}-10^{-2}$ G strength to significantly larger values. Despite the success of the local dynamo simulations, the discussion on the maximum magnetic field strength obtained in such simulations is still on-going.  \citet{Rempel2014} has shown that high resolution (i.e. high Reynolds numbers) alone is not sufficient to reach the magnetic field values suggested by observations. Only if the simulation is allowed to advect magnetic field  from the bottom boundary \citep{Rempel2014}, the obtained mean magnetic field strength reaches values similar to observations based on the Zeeman effect by  \citet{Danilovic2010b}, but still falls short by a factor of two in comparison to the Hanle effect measurements by \citet{TrujilloBueno2004} and  \citet{Shchukina+TrujilloBueno2011}. 
In addition, there are still discussions on observational side about the fraction of the observed magnetic fields in the quiet Sun produced by the global or local dynamo. \citet{Stenflo2013} suggested that local dynamo plays no significant role at any of the spatially resolved scales. Therefore, the last word about the properties and origin of the quiet solar magnetism has not been said yet.

An important but rarely mentioned aspect of the local dynamo theory is the origin of the seed field. All existing simulations of the local solar dynamo use an approach based on ideal MHD equations and assume an initial seed field without addressing its origin. In ideal MHD, the induction equation tells us that the magnetic field cannot be created neither destroyed. Diffusion does not help as it is unable to act on zero field. The initial seed might be of primordial origin. 
While a solar primordial field could be rather strong according to the evidences from meteorites \citep{Ip1984},  it is considered that this field has vanished long time ago diffusing back into space. Remnants of primordial field not larger than \hbox{30 G} in the solar core \citep{Boruta1996}, and less than \hbox{1 G} at the bottom of the convection zone \citep{Gough+McIntyre1998} may exist, completely separate from the surface.  Another possibility is that the magnetic field from active regions is transported to the quiet Sun, providing seeds and perturbing the quiet Sun magnetism. This however would make the properties of the internetwork field cycle dependent, which seems not to be the case. 

A different alternative was suggested by \citet{Biermann1950}, who has shown that the field can be generated by the battery effect.  Any random increase in the electron pressure in a small region leads to a flux of electrons leaving this region. This produces a charge imbalance and an electric field strong enough to resist the electron motion. If the electric field is not curl free, a current starts propagating in the plasma leading to a magnetic field that is variable in time, which in turn produces an electric field, closing the loop \citep{Biermann1950, Kulsrud2008}. The Biermann battery effect is always present
in any plasma and it naturally provides a cycle independent seed for local dynamo.

While the battery effect has been studied widely in the context of galactic dynamos,  no attempts have been made to apply this theory in the context of local turbulent dynamo acting in the subsurface layers of the Sun.  It has not been evaluated if the battery effect acting in the upper layers of the solar convection zone is efficient enough to establish a seed field of sufficient strength to allow for the subsequent amplification by the local dynamo, how the spatial distribution  of this field is, and whether its continuous generation has any influence on the working of the local dynamo. With the present article we attempt to fill this gap.


\begin{figure}
\begin{center}
\includegraphics[width = 8.5cm]{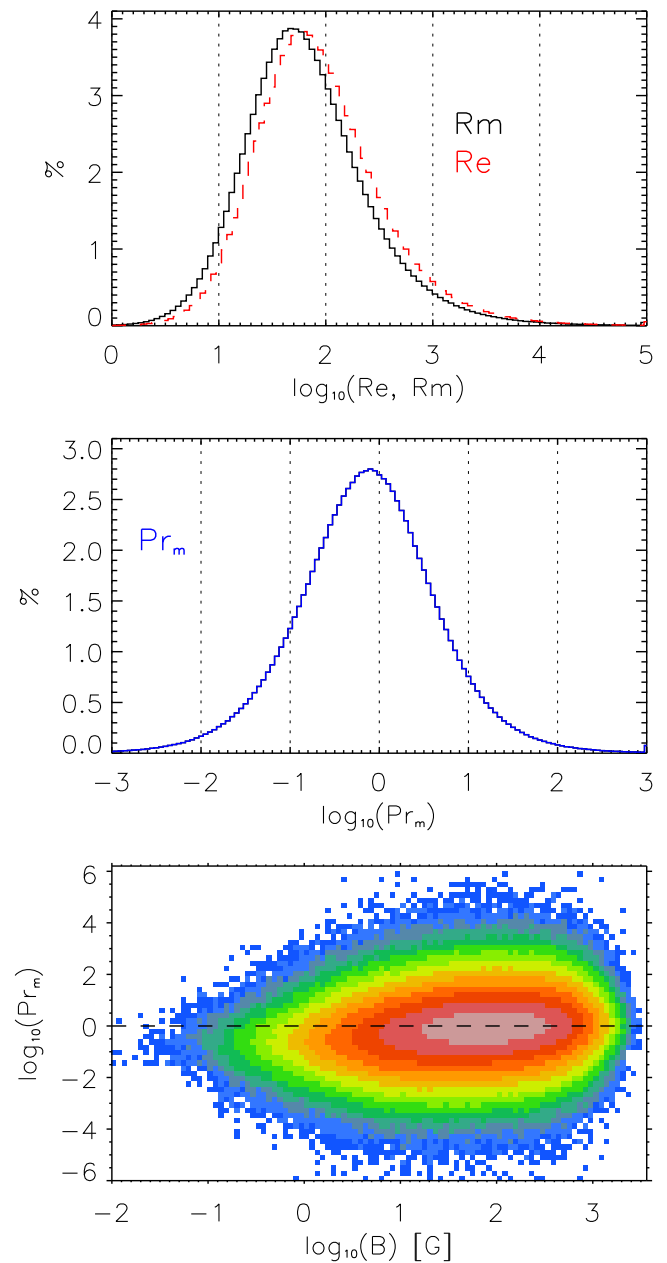}
\end{center}
\caption{Top: histogram of magnetic and hydrodynamic Reynolds numbers. Middle: histogram of the magnetic Prandtl number. Bottom: density plot of the magnetic Prandtl number as a function of the modulus of the magnetic field. Colors from blue to red indicate progressively larger number of data points. The data points include the complete simulation domain of the B20 run.}\label{fig:reynolds}
\end{figure}

\section{Methods}

The simulations presented in this paper were done with the {\sc Mancha3D} code \citep{Khomenko+Collados2006, Felipe+etal2010} that solves the equations of non-ideal magnetohydrodynamics together with a realistic equation of state and non-grey radiative transfer \citep{Khomenko+Collados2012b, Vitas2016}. {\sc  Mancha3D} uses hyper-diffusion algorithms and a Cartesian grid with sixth-order spatial discretization and an explicit fourth-order Runge-Kutta scheme to advance the solution in time. The code is fully MPI-parallelized and allows full arbitrary 3D domain decomposition.  {\sc Mancha3D} solves the radiative transfer equation using the short characteristic method. The code uses a realistic equation of state in thermodynamical equilibrium, which is precomputed and stored in lookup tables for faster access and evaluation. The equation of state includes the ionization equilibrium solution and provides the electron density needed for the computation of the non-ideal terms.

The current version of the code includes the solution of the generalized induction equation (as a consequence of the generalized Ohm's law), including ambipolar, Hall and battery terms. In the present paper we only include the battery term in our model.

We performed a numerical experiment similar to \cite{Vogler+Schussler2007}, and \cite{Rempel2014} but with strictly zero seed field. Instead, the seed was provided by the battery term in the generalized induction equation as explained above. When the simulated hydrodynamical convection fully develops and reaches a stationary regime at solar radiative flux, we switch on the battery term and continue the simulation for several solar hours more, depending on the setup. We used an open bottom boundary condition with mass and entropy controls that ensures that the model has the correct value of solar radiative flux. The bottom boundary has zero magnetic field inflow. This is done by setting magnetic field vertical at the boundary (symmetric $B_z$ and antisymmetric $B_x$ and $B_y$ values in the ghost cells). The top boundary is closed for mass flows. The simulation domain covers $5.8\times5.8\times1.6$ Mm$^3$, with about 600 km being above the visible solar surface. The grid size was of 20 km horizontally and 14 km  vertically.

Table 1 summarizes the parameters used in the four simulation runs performed: two using the battery term (B20 and B20dif) and the other two (R20 and R20dif) using the same numerical setup but with an artificial seed field, of the same kind as in \cite{Vogler+Schussler2007, PietarilaGraham2010, Rempel2014,  Kitiashvili2015}, with an amplitude rms value of $10^{-6}$ G. The particular realization of the random field is not relevant, as was shown by \citet{Kitiashvili2015}. We used field that is constant in the vertical direction and random in the horizontal direction with a white noise realization on the scale of a pixel of our simulation box.

%
%
In the cases B20 and R20, we set both constant diffusion  ($\eta_{\rm c}$) and hyper-diffusion ($\eta_{\rm hyp}$) to zero in the induction equation, so that the diffusion of the magnetic field is caused by the discretization scheme alone. The rest of the MHD equations (continuity, momentum, and energy equations) are subject to a minimum amount of hyper-diffusion, necessary to numerically stabilize the simulations. The runs B20dif and R20dif were done with constant diffusion and hyper-diffusion for $B$ different from zero (the average value over the box is given in the Table for reference).

\begin{figure*}
\begin{center}
\includegraphics[width = 16cm]{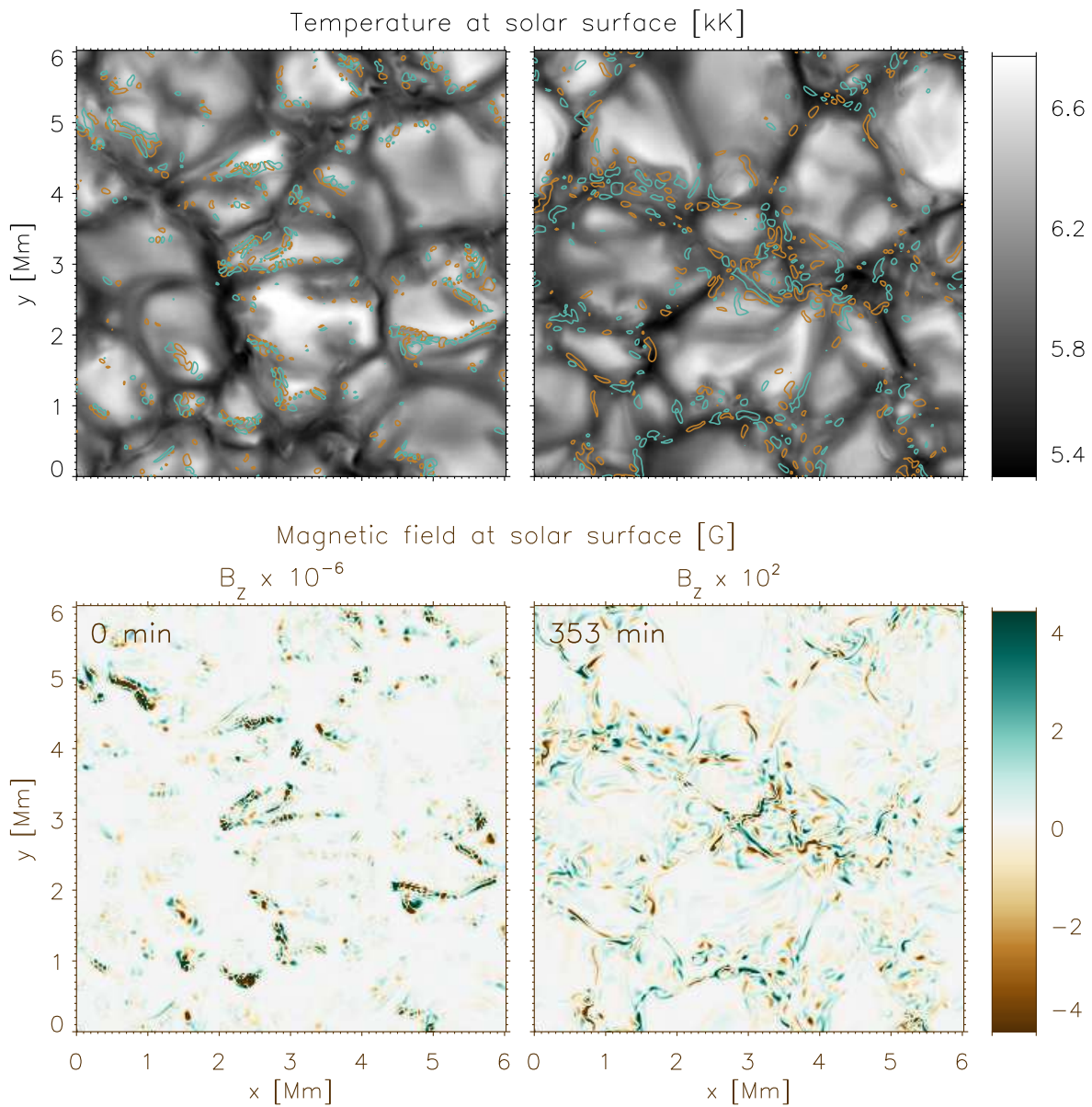}
\end{center}
\caption{Snapshots of the temperature (upper panels) and of the vertical magnetic field component (bottom panels) at $\tau_5=1$ in the simulations at two distinct time moments in time. The snapshots on  the left are taken 10 seconds after the start of the action of the battery; the color scale for the magnetic field strength saturates at $\pm 4\times 10^{-6}$ G. Yellow and orange contours superposed on the temperature map mark the locations with $\pm 1\times 10^{-6}$ G vertical field strength. The snapshots on the right are taken after the dynamo has reached the stationary phase. Here, the colors saturate at $\pm 400$ G. The contours superposed on the temperature map mark locations with $ \pm 1\times 10^{2}$  G field strength. }
\label{fig:snap}
\end{figure*}

\subsection{Reynolds and magnetic Prandtl numbers of the simulations}

Due to numerical reasons, the regime of Reynolds, $Re$ and $Rm$,  and magnetic  Prandtl, $Pr_m$, numbers reached in simulations of magneto-convection and dynamo is different from that of the solar case \citep[see, e.g.,][]{Kaplya2011}. This is so both for the simulations reported here and for those reported previously in the literature \citep{Vogler+Schussler2007, PietarilaGraham2010, Rempel2014,  Kitiashvili2015}. In the Sun, hydrodynamic Reynolds numbers reach values about $Re \sim10^9$ in the photosphere \citep{Komm1991}, while magnetic Prandtl number, $Pr_m=Rm/Re$ vary from about $10^{-2}$ at the bottom of the convection zone to $10^{-5}$ at the surface \citep{Rieutord+Rincon2010}. 

The efficiency of the dynamo depends on the magnetic Prandtl number in a complex manner \citep{Thaler+Spruit2015}.  The magnetic field can be relatively easy amplified by the dynamo for large magnetic  Prandtl numbers, i.e., when the magnetic diffusion is smaller than viscosity \citep[see][]{Brandenburg1996, Schekochihin2004b}. The amplification is still possible for small $Pr_m$ values, at least theoretically, as follows from the recent discussion in \citet{Schekochihin2004a, Schekochihin2007, Tobias2011}.

The evaluation of these characteristic numbers from magneto-convection simulations is not a straightforward task \citep[e.g.,][]{Porter+Woodward1994} since many scales are involved. The few papers where those numbers are reported provide typical values of \hbox{Re $\sim10^2-10^3$}  and \hbox{$Pr_m \sim 10^{-1}-10^{0}$} \citep{Vogler+Schussler2007, Kitiashvili2015}. Simulations using hyper-diffusion algorithms for numerical stability, like the ones reported here, employ hyper-diffusion terms to damp fluctuations at small scales. The values of the hyper-diffusion coefficients are calculated depending on the scale of variation of a given parameter. This strategy leaves us with artificial magnetic diffusivity and viscosity coefficients varying in space. Similarly, no single velocity and spatial scale can be assigned to the flows, since the parameters are strongly spatially dependent. 

Here we have evaluated the Reynolds and magnetic Prandtl numbers using their canonical definition:
\begin{eqnarray}
Re&=&\frac{|\rho(\vec{u}\vec{\nabla})\vec{u}|}{|\vec{\nabla}\hat{\tau} |} , \\ \nonumber
Rm&=&\frac{|\vec{\nabla}\times(\vec{u}\times\vec{B})|}{|\vec{\nabla}\times(\eta\vec{\nabla}\times\vec{B})|} ,\\ \nonumber
Pr_m&=&\frac{Rm}{Re},
\end{eqnarray}
where $\hat{\tau}$ is the stress tensor:
\begin{equation}
\tau_{ij}=\frac{\rho}{2}\left( \nu_i\frac{\partial u_j}{\partial x_i} + \nu_j\frac{\partial u_i}{\partial x_j} \right).
\end{equation}

The magnetic diffusivity $\eta$ and the viscosity $\nu$ are computed a posteriori from the simulation snapshots using values of $\eta_c$, $\nu_c$, $\eta_{\rm hyp}$ and $\nu_{\rm hyp}$ coefficients that are evaluated in the identical way as in the code \citep{Felipe+etal2010}. Additionally, the intrinsic diffusion of the numerical scheme should be added to the $\eta$ and $\nu$ coefficients. The runs B20 and R20 have the coefficients $\eta_c$ and $\eta_{\rm hyp}$ set to zero. In these two cases, the only diffusion acting on the magnetic field is the diffusion caused by the numerical scheme.  In order to estimate the diffusion properties of the numerical scheme, we assumed a constant value of $\eta_{\rm scheme}$ and made a linear regression between the corresponding terms in the induction equation:
\begin{equation}
\vec{y}=\eta_{\rm scheme}\vec{x}+c,
\end{equation}
where
\begin{eqnarray}
\vec{x}&=&\Delta\vec{B}, \\ \nonumber
\vec{y}&=&\partial\vec{B}/\partial t - \vec{\nabla}\times (\vec{u}\times\vec{B}) .
\end{eqnarray}
where $\Delta$ is Laplacian operator. The temporal derivative is calculated for snapshots saved with a cadence of 0.4 sec. The regression was made separately for all three components of $\vec{B}$, giving similar results for $\eta_{\rm scheme}$. The resulting value was calculated as a modulus over three components providing  $\eta_{\rm scheme}=2.5 \times 10^6$ m$^2$/s. 

The Reynolds and magnetic Prandtl numbers calculated as above are presented in Figure \ref{fig:reynolds}.  The distributions are rather broad, with average values of $Re\approx 940$ and $Rm\approx 680$. The histogram of the magnetic Prandtl number peaks at values slightly lower than 1. However, no definite dependence between the value of $Pr_m$ and the magnetic field strength is found (bottom panel of Fig.\ref{fig:reynolds}). In the case of  the runs R20dif and B20dif, $Rm\approx 90$, i.e. about an order of magnitude smaller than for the runs R20 and B20. 

\begin{figure}
\begin{center}
\includegraphics[width = 9cm]{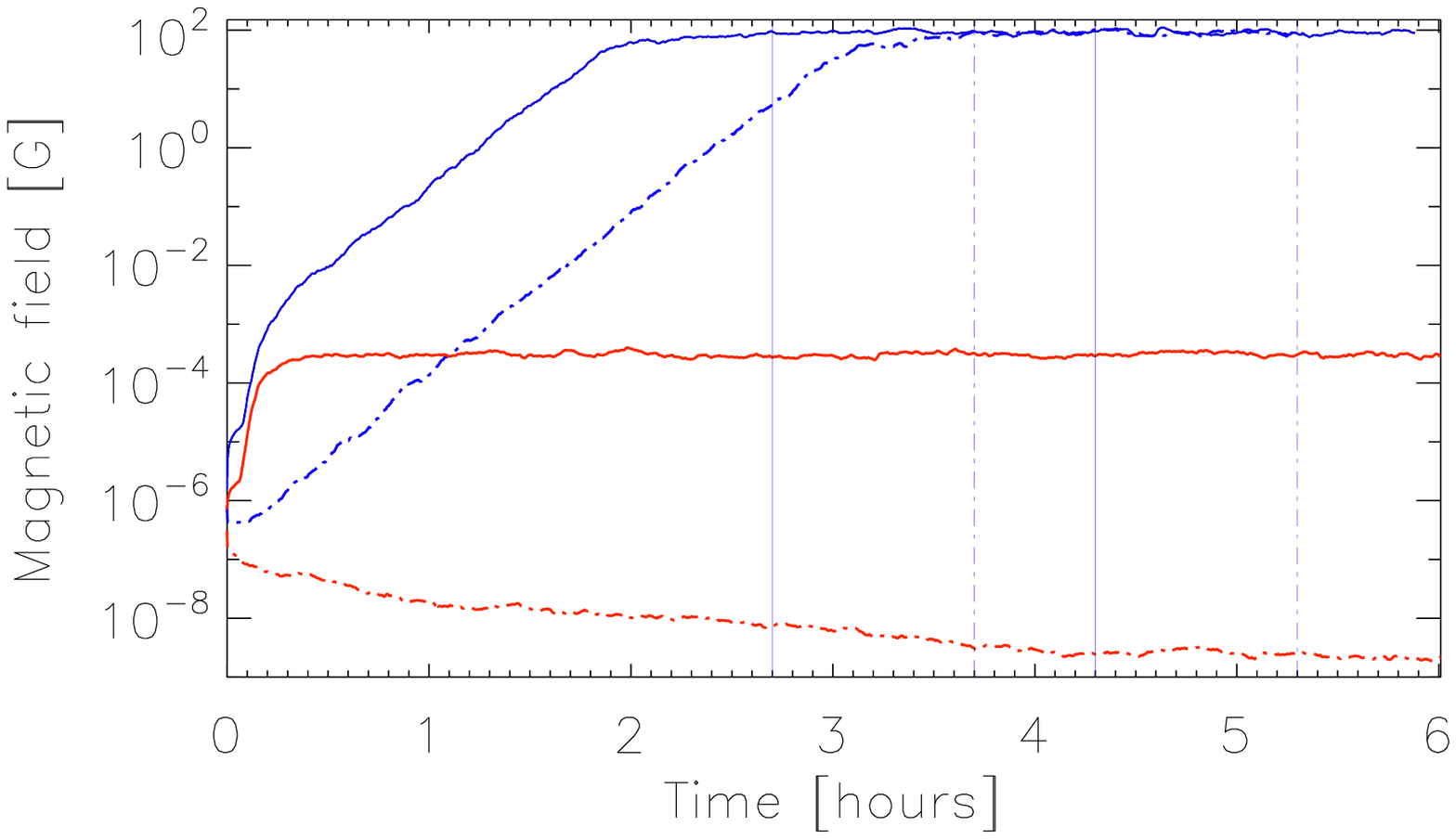}
\includegraphics[width = 9cm]{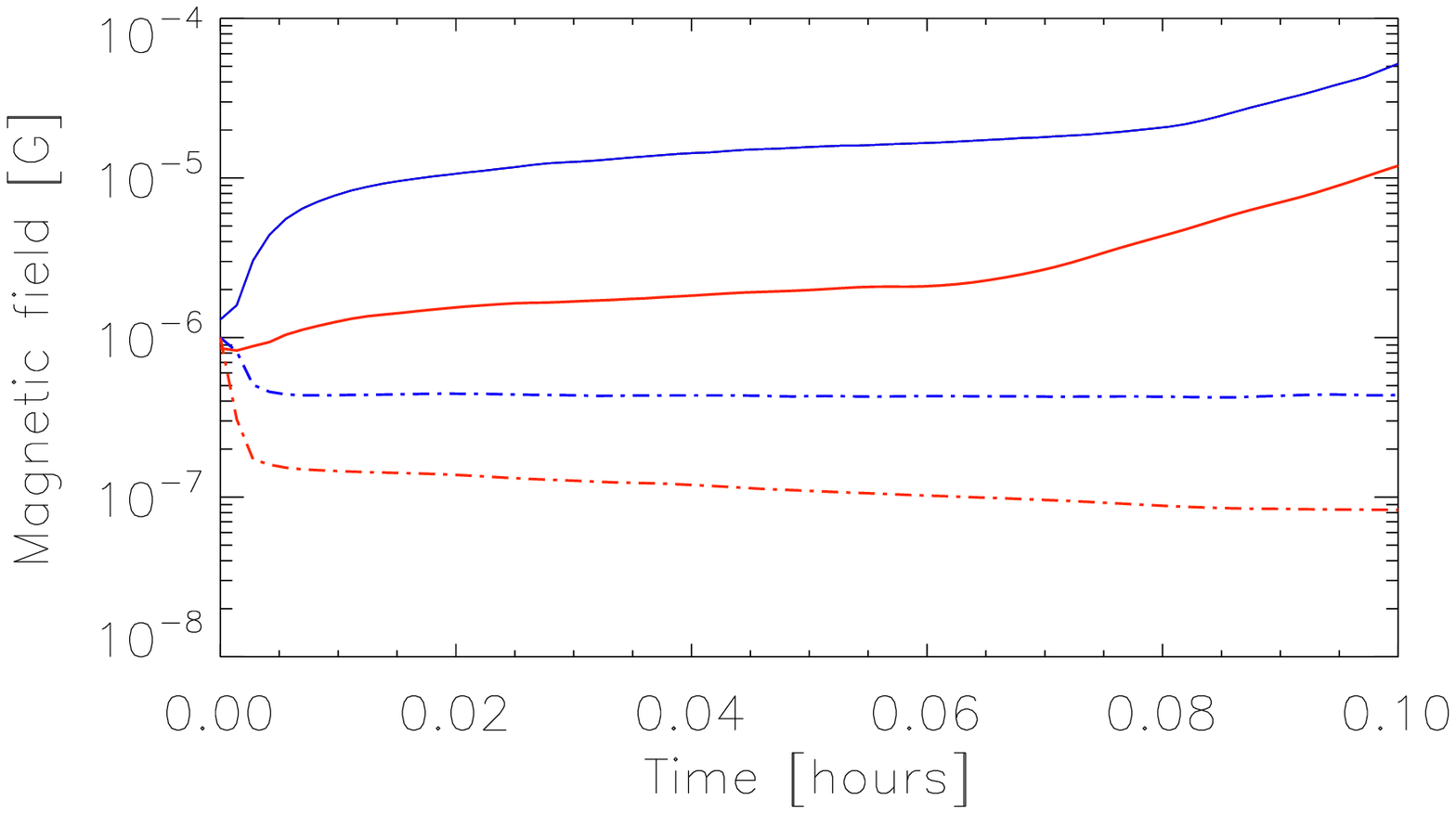}
\end{center}
\caption{Top panel: mean value of the modulus of the magnetic field vector on the surface with optical depth $\tau_5=1$ as a function of time after the start of the action of the battery or the introduction of the random seed. Solid blue: B20; dashed blue: R20; solid red: B20dif; dashed red: R20dif. Bottom panel: detail of the above figure showing only the first 0.1 hours of the simulations. }\label{fig:rate}
\end{figure}

\begin{center}
\begin{table}
\begin{center}
\caption{Parameters of the simulation runs} 
\begin{tabular}{ccccc}
\hline ID &  Seed & $\eta_{\rm c}$ [m$^2$/s] & $\eta_{\rm hyp}$ [m$^2$/s]  & Seed  [G] \\ \hline
B20 & battery & 0  & 0 & $-$ \\
R20 & random & 0 & 0 & $10^{-6}$ \\
\hline
B20dif & battery & $8\times10^6$ & $2\times10^5$  & $-$ \\
R20dif & random & $8\times10^6$ & $2\times10^5$ & $10^{-6}$ \\
\hline
\end{tabular}
\end{center}
\end{table}
\end{center}

\section{Battery effect}

The generalized Ohm's law, applicable to the multi-component solar plasma, can be written in the following form \citep{Pandey+Wardle2008, Zaqarashvili2011, Leake+etal2014, Khomenko+etal2014b, Ballester2017}:

\begin{eqnarray} \label{eq:ohm}
\vec{E}^*= \eta\vec{J} +   \eta_A\frac{[(\vec{J} \times \vec{B}) \times \vec{B}]}{|B|^2} + \frac{[\vec{J} \times \vec{B}] }{en_e} - \frac{\vec{\nabla}{p_e}}{en_e} ,
\end{eqnarray}
where $\vec{E}^*=[\vec{E}+\vec{u}\times\vec{B}]$ is the electric field, and the coefficients $\eta$ and $\eta_A$ are given by:
\begin{equation} 
\eta_A=\frac{\xi_n^2 |B|^2}{(\rho_e\nu_{en} + \rho_i\nu_{in})}; \,\,\, \eta=\frac{\rho_e(\nu_{ei}+\nu_{en})}{(en_e)^2}.
\end{equation}
In these equations, $\nu_{in}$, $\nu_{ei}$ and $\nu_{en}$ are the collisional frequencies between electrons, ions and neutrals; $\xi_n=\rho_n/\rho$ is the neutral fraction and  $\rho_i$, $\rho_e$ and $\rho_n$ are mass densities. The terms present in the Ohm's law reflect the ohmic and ambipolar diffusion effects, and the Hall and battery effects. The ambipolar term is only present when the plasma contains a neutral component (non-zero $\xi_n$). The other three terms are always present in any plasma irrespectively of the degree of ionization. Frequently, all these terms are dropped assuming they are relatively small. Here we are only interested in the battery effect and therefore only retain the corresponding term.

The magnetic induction equation with only the battery term can be expressed as:
\begin{equation}
\frac{\partial\vec{B}}{\partial t} =  \vec{\nabla}\times \left (\vec{u}\times\vec{B} +\frac{\vec{\nabla}{p_e}}{en_e}\right).
\end{equation}
It can be seen that the battery term is independent of the magnetic field and acts as a source term in the induction equation. Due to its smallness, the numerical treatment does not represent any difficulty, and no changes in the usual integration scheme (as operation splitting) is required. 

The above induction equation can be rewritten in the same form as the equation for the evolution of 
the vorticity, see \citet{Kulsrud2008}:
\begin{eqnarray}
\frac{\partial\vec{B}}{\partial t}  &=&  \vec{\nabla}\times (\vec{u}\times\vec{B}) + \frac{\vec{\nabla}p \times \vec{\nabla}\rho}{\rho^2}\frac{\mu}{e}, \\ \nonumber
\frac{\partial\vec{\omega}}{\partial t}  &=&  \vec{\nabla}\times (\vec{u}\times\vec{\omega}) - \frac{\vec{\nabla}p \times \vec{\nabla}\rho}{\rho^2}. \\ \nonumber
\end{eqnarray}
In this derivation we have assumed a constant mean molecular weight, $\mu$, and a constant ratio between ion and total number densities (this ratio is equal to 0.5 for a fully ionized plasma and is above 0.5 for a partially ionized plasma).

The comparison between the induction and vorticity equations provides a way of evaluating the strength of the initial magnetic field produced by the battery effect in a unit time  from the simple relation:
\begin{equation} \label{eq:bini}
\frac{e |\vec{B}|}{\mu}\sim |\vec{\nabla}\times\vec{v}| .
\end{equation}
Once the field is generated by the battery effect, it is amplified by the flow via the dynamo term.

\begin{figure*}
\begin{center}
\includegraphics[width = 9cm]{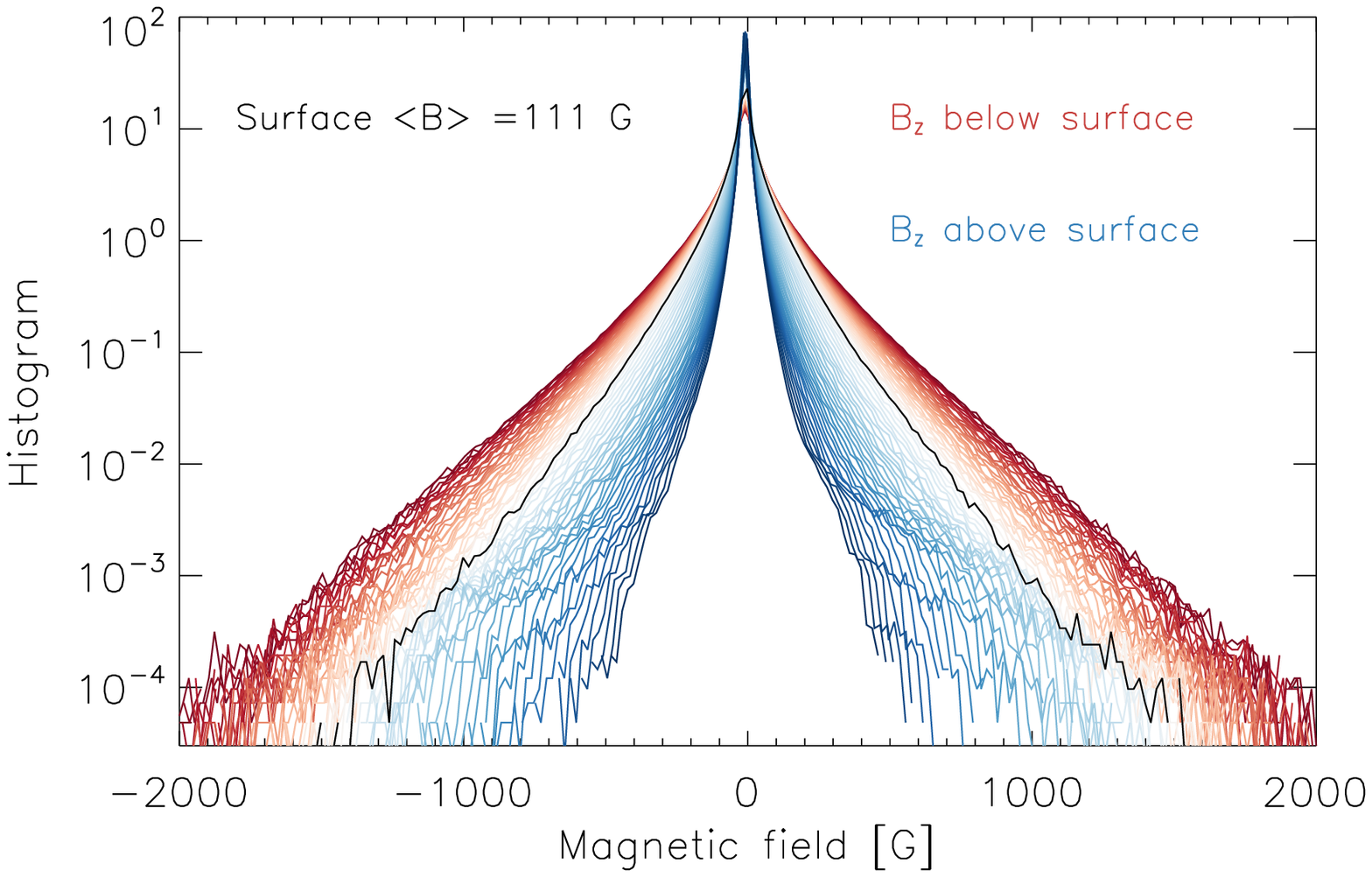}
\includegraphics[width = 9cm]{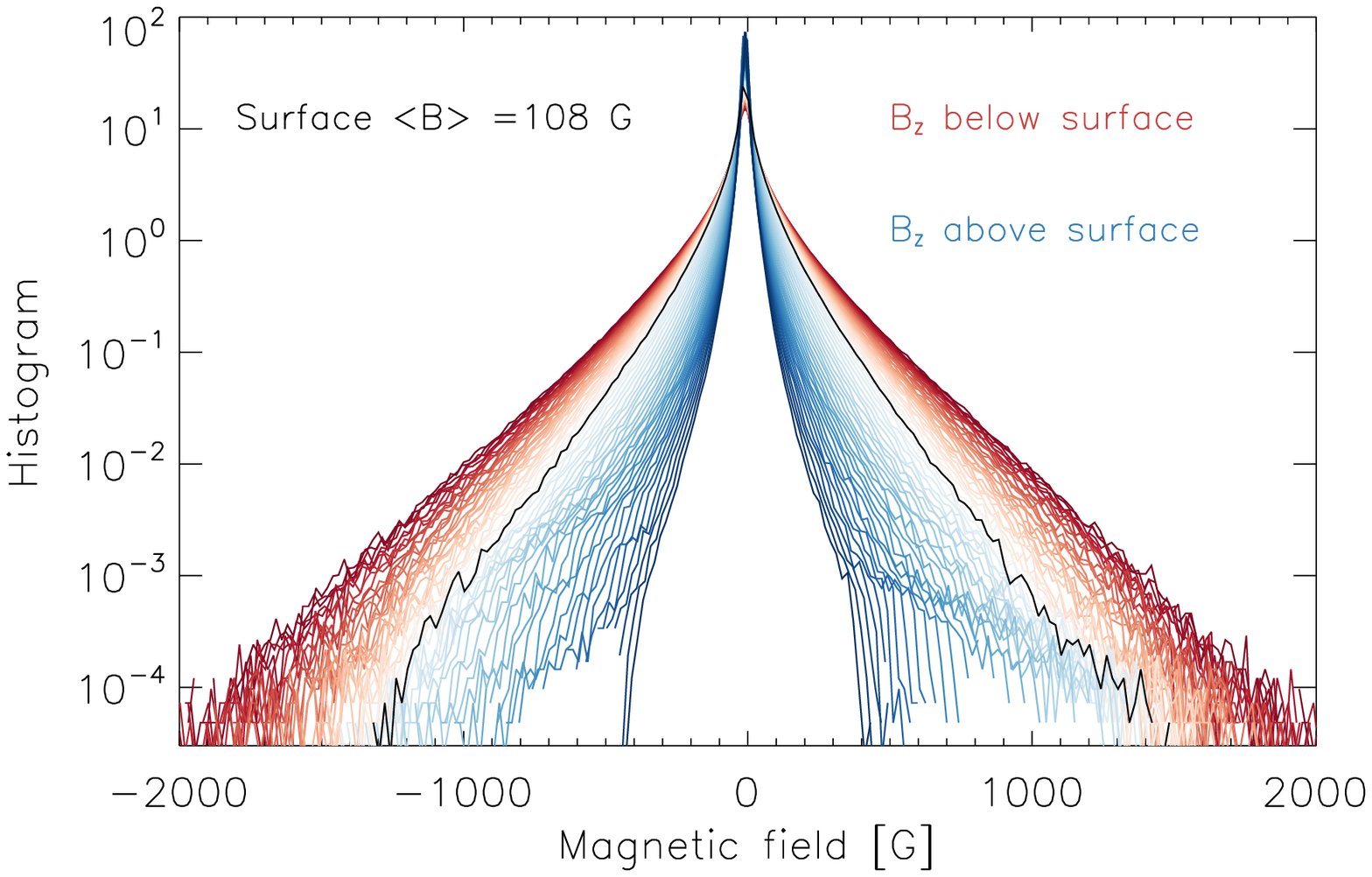}
\end{center}
\caption{Left panel: histograms of the vertical magnetic field component, $B_Z$,  in the stationary dynamo phase of the run B20. The histograms in red colors refer to heights below the solar surface from $-420$ to 0 km; the histograms in blue colors refer to heights above the solar surface from 0 to 420 km.  The black curve corresponds to the histogram of $B_Z$ at the solar surface $\tau_5$ = 1. Right panel: same, for the run R20.}\label{fig:hist}
\end{figure*}

\begin{figure*}
\begin{center}
\includegraphics[width = 16cm]{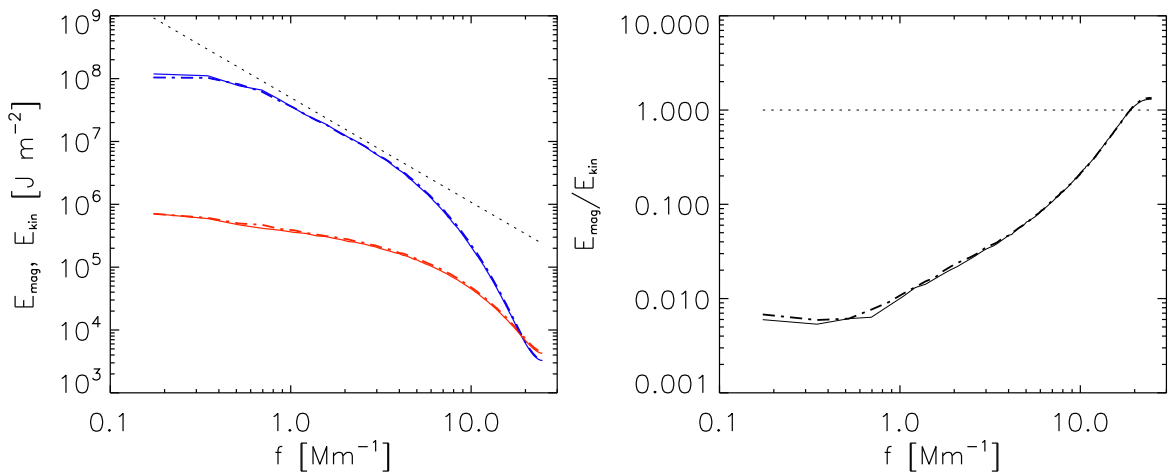}
\end{center}
\caption{Left panel: spatial power spectra of the kinetic (blue) and magnetic (red) energies in the B20 (solid lines) and R20 (dashed lines) runs at $\tau_5=1$. The black dotted line is the -5/3 Kolmogorov power law. Right panel: ratio of magnetic to kinetic power spectra for the B20 (solid line) and R20 (dashed line) runs. The dotted line indicates the equipartition level between magnetic and kinetic energies.}\label{fig:scales}
\end{figure*}

\begin{figure*}
\begin{center}
\includegraphics[width = 16cm]{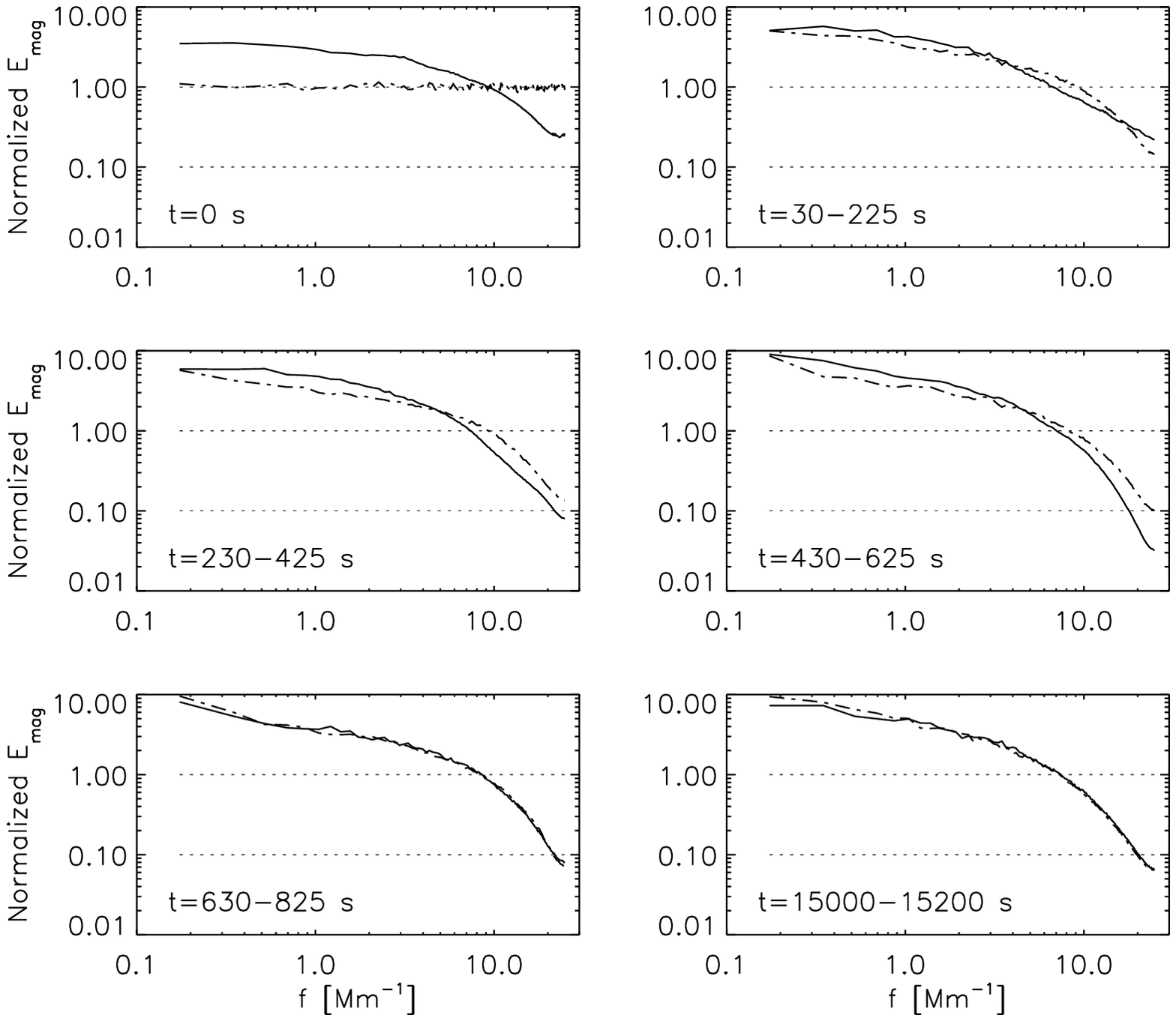}
\end{center}
\caption{Time evolution of the magnetic energy power spectra in the B20 (solid lines) and R20 (dashed lines) runs. Time is indicated at each panel. In order to make the comparison, the spectra have been normalized in such a way that the total magnetic energy is constant at each time moment.}\label{fig:scales_time}
\end{figure*}

\section{Results of the simulations}

Figure \ref{fig:snap} shows snapshots of the temperature and the vertical component of magnetic field right after the battery was introduced (left panels) and when the dynamo has reached the stationary stage (right panels), corresponding to the run B20. The magnetic field is not homogeneously generated and appears at locations with strong electron pressure gradients, which coincide with the borders of the granular convection cells. The field distribution is completely different in the initial and stationary stages, since in the latter almost all strong fields have been advected by the flow into the intergranular downflowing lanes (upper right panel of Figure \ref{fig:snap}).

The top panel of figure \ref{fig:rate} shows the mean magnetic field strength at the height corresponding to the solar surface (defined to be the height where the optical depth is equal to one) as a function of time, for all four runs. The bottom panel of this figure shows a detail of the evolution for the first 0.1 hours. It can be seen that the initial field produced in the first seconds of the simulation by the battery seed is of the order of $10^{-6}$ G, i.e. three orders of magnitude smaller than the artificially seeded  field by \cite{Vogler+Schussler2007}, and \citet{Rempel2014}.  A similar number ($10^{-6}$ G) is also obtained after an order of magnitude evaluation from Equation \ref{eq:bini} using a typical velocity and spatial scale representative of granules for the evaluation of the vorticity. 
Figure \ref{fig:rate} demonstrates that this field is rapidly amplified going through various stages (blue solid curve). In the initial stage covering about 20 min, the battery effect dominates the dynamo action producing a linear growth till the field reaches about $10^{-3}$ G (see bottom panel). After that time, both battery and dynamo terms contribute to the field growth. The battery term is continuously operating (at a mean field production rate of a few $10^{-8}$ G/s) because of the continuous action of convection in the solar atmosphere, making it efficient to seed the advective term. 

The saturation field obtained in the stationary phase is due to the balance between the kinematic dynamo action and diffusion. In both simulations with random seed, R20, and with battery seed, B20, we reach a field strength in the stationary phase of about $10^2$ G. This number is very close to the values retrieved from solar observations  \cite{TrujilloBueno2004, SanchezAlmeida+MartinezGonzales2011}.  The histograms of the magnetic field strength presented in Figure \ref{fig:hist} are also very similar to those derived from solar observations of the quiet Sun magnetic fields. The comparison between the histograms in the R20 and B20 cases reveals that they are undistinguishable.

We can compare our results with those from local dynamo simulations initiated with an artificial random field as in \cite{Vogler+Schussler2007, Rempel2014,  Kitiashvili2015}. 
The growth curve in the R20 run (blue dashed line in Fig.\ref{fig:rate}) shows that it takes a somewhat longer time to arrive to the same saturation level of magnetic field strength with an initial random distribution, compared to the case in which the field is naturally provided by the battery seed,  when both start from the same strength of the seed field. This difference in the growth times is not so important given that the local dynamo operates on relatively short time scales in both cases. Both for the random and for the battery seeds, similar saturation levels are obtained.  

Figure \ref{fig:scales} shows the spatial power spectra of the kinetic and magnetic energies of the R20 and B20 simulations at the stationary regime, corresponding to $\tau_5=1$. The spectra are essentially the same, indicating once again that the final state reached in the simulations is independent of the initial seed. The Kolmogorov-like power law exists in the kinetic energy in the central part of the spectrum, while no such dependence is seen in magnetic energy. These spectra are similar to those given in \cite{Rempel2014} for the lower resolution runs. The super-equipartition between kinetic and magnetic energies starts for scales smaller than 60 km. The time evolution of the magnetic energy power spectra  of the R20 and B20 runs are compared in Figure \ref{fig:scales_time}. One can observe that while there is no preferred scale in the initial R20 field distribution (in accordance with its random nature), the battery term generates the field with a spatial power spectrum whose shape is qualitatively similar to the one at the saturated state. The subsequent time evolution of the spectra shows that the flows in the R20 run quickly change the shape of the $E_{\rm mag}$ distribution, making the R20 and B20 similar after about 10 min of the simulations. However, while the initial B20 field distribution allows its immediate amplification, the R20 one produces initial cancellation and readjustment of the field, leading to initial decrease of the mean B, seen in Figure \ref{fig:rate}.  

The result that the battery effect generates sufficiently strong field with a spatial spectrum that facilitates its quick amplification could not be obtained a priory from general considerations and implies far-reaching conclusions. It is possible that the local dynamo that is believed to generate the quiet Sun magnetism works completely self-consistently without the need of an external seed field. The battery term is always present in a plasma under the sole condition that gradients of electron pressure and temperature are not parallel to each other, which is very easy to fulfill. Irrespectively from any other possible mechanisms, the battery effect will always generate seeds. In particular, the outer solar convection is very effective in generating such seeds through the battery effect, implying that the solar plasma at the surface will always be magnetized. Our results demonstrate that quiet Sun magnetic fields can be present independently from global magnetism, existence of active regions or  fossil field from where the initial seed for the local dynamo might come. It also implies that the quiet Sun component must have  similar properties at all latitudes, as far as convection can be considered isotropic and if we only take into consideration the seed by the battery term. 

Both simulations R20 and B20 were done at the highest possible Reynolds numbers for a given resolution by setting the magnetic hyper-diffusion coefficient to zero. If magnetic hyper-diffusion is present (cases R20dif and B20dif), the growth curves significantly change, see red curves in Figure \ref{fig:rate}. In the case of battery seed, B20dif, the growth and saturation are still present, but the saturation happens at significantly lower field strengths than in the case B20, determined by numerical dissipation. In the random seed case, no dynamo is possible anymore. The latter result is consistent with previous findings. Although a much lower saturation level is reached in the B20dif simulation than in the B20 simulation, it nevertheless shows a fundamental difference with the random seed case, R20dif. The continuous seeding by battery is able to resist the dissipation of the field due to diffusivity at small scales, and saturation is possible even for small Reynolds numbers. In simulations with a different amplitude of the magnetic hyper-diffusion coefficient we observe that the saturation happens at a different magnetization level, but the saturation regime is always reached if the seed is produced by the battery, unlike to the case of random seeding. This conclusion must be further explored in future studies.

Following the estimates from Equation \ref{eq:bini}, we expect that the properties of the battery-generated seed field will not change significantly with the resolution of the simulations. Nevertheless, the improved resolution in future models with battery seed and subsequent dynamo action may allow for lower dissipation rates leading to an even larger magnetization.

\section{Discussion and Conclusions}

Observational properties and the origin of the magnetic field in the internetwork of the quiet Sun are under vivid discussion in the literature. The magnetic field in those quiet regions is believed to be generated by local dynamo action from some initial seed field. Such a local dynamo is thought to exist independently of the solar cycle. To start the action of the local dynamo in the approximation of ideal MHD, an initial seed field is required since the ideal induction equation can not produce magnetic field from zero. All existing models of solar local dynamo start with an arbitrary random seed field with an amplitude in the range of 10$^{-6}$ - 10$^{-2}$  G \citep{Rempel2014, Kitiashvili2015}. It has been stated \citep{Kitiashvili2015} that the seed amplitude does not influence the final magnetization when the dynamo is saturated. However, not any seed must serve since its amplitude and spatial structure must be adequate to allow efficient dynamo amplification, and to resist damping by Ohmic diffusion at small spatial scales. The seeds from fossil field of the Sun possibly have too low amplitude \citep{Boruta1996, Gough+McIntyre1998}, and their spatial distribution is known only very approximately. While the seeds from active regions recycling may cause cycle dependence of the quiet Sun magnetism, which has not been clarified observationally. 


We have shown that a seed field of sufficient amount can be naturally created by the battery effect alone, arising from local fluctuations of the electron pressure due to convective motions. The battery term contributes to the induction equation as an additional forcing term and its action is continuously present. Although the term itself is small, it cannot be said based on intuition or simple reasoning that it does not affect the statistical behavior of the turbulent small-scale dynamo in the Sun. The effects of a forcing term in the induction equation are subject of detailed studies even in much more idealized setups, e.g. by \citet{Krstulovic2014}.

Our results demonstrate that the magnetization of the quiet Sun in the solar outer convection layers may exist independently of any external source and at the same level as indicated by observations, under the condition that the local dynamo is at work. In other words, even if the Sun had no global cycle, it would nevertheless have local magnetization in the convective layers, in agreement with the X-ray behavior of late-type fully-convective stars \citep{Wright+Drake2016}. 

An important consequence of our results is that the Biermann battery term provides a lower bound on the seed field in the solar convective layers. This way it sets a relevant upper limit to the timescale (of about an hour) to reach the observed level of magnetization of the quiet Sun. It can be stated that everywhere on the surface the Sun there would be detectable, close to saturated, local dynamo fields within an hour or so after plasma has newly emerged, even if the emerged plasma has extremely low field strengths.


{\bf Acknowledgements.} This work was supported by the Spanish Ministry of Science through the project AYA2014-55078-P and by the European Research Council in the frame of the FP7 Specific Program IDEAS through the Starting Grant ERC-2011-StG 277829-SPIA. We acknowledge PRACE for awarding us access to resource MareNostrum based in Barcelona/Spain. We thank Robert Cameron and Manfred Sch\"ussler for constructive discussions and suggestions that significantly improved our manuscript, and Vyacheslav (Slava) Lukin for the idea of evaluation of diffusion properties for the numerical scheme in Section 2.1. 

\input{ms_lit.bbl}

\end{document}

%% file: ms_lit.bbl
\providecommand{\noopsort}[1]{}\providecommand{\singleletter}[1]{#1}%